\documentclass[a4,twocolumn]{revtex4}
\usepackage[utf8]{inputenc}
\usepackage[pdftex]{graphicx}
\usepackage{amsmath,amssymb}
\usepackage{dcolumn}
\usepackage{bm} 
\usepackage{lipsum}
\usepackage{lineno}

\newcommand{\ave}[1]{\ensuremath{\langle#1\rangle}}
\newcommand{\dd}[2]{\frac{\textrm{d} #1}{\textrm{d} #2}}

\newcommand{\pp}[2]{\frac{\partial  #1}{\partial #2}}
\newcommand{\ppt}[2]{\frac{\partial ^2  #1}{\partial #2 ^2}}

\newcommand{\KL}[1]{D_\mathrm{KL}({#1})}
\def\d#1{\textrm{d}#1}

\begin{document}
\title{Stochastic thermodynamic limit on \textit{E. coli} adaptation by Information geometric approach}

\author{Keita Ashida}
\affiliation{Department of Biosciences and Informatics, Keio University, 3-14-1, Hiyoshi, Kohoku-ku, Yokohama, Kanagawa, 223-8522, Japan}
\author{Kotaro Oka}
\affiliation{Department of Biosciences and Informatics, Keio University, 3-14-1, Hiyoshi, Kohoku-ku, Yokohama, Kanagawa, 223-8522, Japan}
\affiliation{Graduate Institute of Medicine, College of Medicine, Kaohsiung Medical University, Taiwan}
\affiliation{Corresponding author: oka@bpni.bio.keio.ac.jp}

\begin{abstract}
 Biological systems process information under noisy environment.
Sensory adaptation model of \textit{E. coli} is suitable for investigation because of its simplicity.
 To understand the adaptation processing quantitatively, stochastic thermodynamic approach has been attempted.
 Information processing can be assumed as state transition of a system that consists of signal transduction molecules using thermodynamic approach, and  efficiency can be measured as thermodynamic cost.
 Recently, using information geometry and stochastic thermodynamics, a relationship between speed of the transition and the thermodynamic cost has been investigated for a chemical reaction model.
 Here, we introduce this approach to sensory adaptation model of \textit{E. coli}, and examined a relationship between adaptation speed and the thermodynamic cost, and efficiency of the adaptation speed.
 For increasing external noise level in stimulation, the efficiency decreased, but the efficiency was highly robust to external stimulation strength.
Moreover, we demonstrated that there is the best noise to achieve the adaptation in the aspect of thermodynamic efficiency.
 Our quantification method provides a framework to understand the adaptation speed and the thermodynamic cost for various biological systems.
\end{abstract}

\maketitle

\section{Introduction}
In biological systems, external information is processed under noisy environment.
Sensory adaptation is one of the important factors for information processings to adjust sensitivity in various strength of stimulation.
To implement the adaptation, negative feedback is widely observed in numerous biological systems \cite{nakatani1991,menini1999,tu2008,hohmann2002,hazelbauer2008,lan2012,shidara2017}.
To understand the process under noise quantitatively, stochastic thermodynamic approach has been used \cite{sartori2011, lan2012, sartori2014, ito2015, ito2017}.
Especially, sensory adaptation of $\textit{Escherichia coli}$ chemotaxis is suitable model to investigate \cite{hazelbauer2008,micali2016,bi2018},
 and thermodynamic cost on the adaptation is well understood \cite{tu2008, sartori2011, lan2012, ito2015}.
In \textit{E. coli} sensory system, an external ligand is received by a receptor named methyl-accepting chemotaxis protein (MCP), and inhibits autophosphorylation of a kinase CheA.
CheA phosphorylation inhibits forward movement \textit{via} CheY activation.
CheA phosphorylation also phosphorylates CheB, and CheB inhibits MCP by methylation of MCP.
This negative feedback, which consists of the kinase activity and methylation level, leads the adaptation to the ligand (Fig. \ref{fig0}).
In previous research \cite{lan2012}, a relationship between energy-speed-accuracy of the adaptation is formulated.
In the research, the accuracy of the adaptation is represented by a difference between initial level and the level after the adaptation of the kinase, and 
the speed of adaptation is demonstrated by time constant of the adaptation model.
The thermodynamic cost can be formulated by these parameters.
There is no way to consider an efficiency of the adaptation speed over time by this relationship because only time constant can be discussed.
Recently, by introducing information geometry to stochastic thermodynamics, a relationship between speed of chemical-state transition and the thermodynamic cost is described under a chemical reaction model \cite{ito2017}.
Moreover, general Langevin equation is investigated by information geometric approach \cite{ito2018}.
This approach enables to argue efficiency of the adaptation speed over time.

Here, we introduce the information geometric and stochastic thermodynamic approach to sensory adaptation model of \textit{E. coli}.
A relationship between speed of the adaptation and thermodynamic cost was investigated by numerical simulation on this adaptation model.
As a result, the efficiency showed robustness for external stimulation strength,
but for increasing external noise level in stimulation, the efficiency decreased.
Moreover, there was an appropriate noise level for efficient adaptation.

\section{Materials and Methods}
\subsection{Sensory adaptation model of \textit{E. coli}}
For mathematical model of \textit{E. coli} adaptation model for chemotaxis, we used coarse-grained coupled Langevin equations described in previous research \cite{tu2008, sartori2011, ito2015} as 
\begin{align}
 \dd{a(t)}{t} &= -\frac{1}{\tau_a}[a(t) - \overline{a(t)}] + \xi_a(t) \notag \\
 &= -\frac{1}{\tau_a}[a(t) - \alpha m(t) + \beta l(t)] + \xi_a(t)
 \label{eq:a}\\
 \dd{m(t)}{t} &= -\frac{1}{\tau_m}a(t) + \xi_m(t)
 \label{eq:m}
\end{align}

where $a(t)$ is a kinase activity, $m(t)$ is a methylation level of a receptor, and $l(t)$ is an external ligand level.
$\overline{a(t)}$ is a stationary value of the kinase activity under the adaptation (Fig. \ref{fig0}).
$\xi_a(t)$ and $\xi_m(t)$ are independent Gaussian white noise which strength is $2T_a$ and $2T_m$, respectively.
This model is coarse-grained models, but validated by experiments \cite{tu2008}.
 From previous research, parameters were determined as follows: $\tau_a = 0.02$, $\tau _m = 0.2$, $\alpha = 2.7$ and $T_a = T_m = 0.005$ \cite{tu2008,tostevin2009,lan2012}.
 We noted that the parameters are dimensionless quantities.
 
\subsection{Numerical simulation}
We performed numerical simulation for calculating a square of an infinitesimal distance $ds^2$, the thermodynamic cost change $\mathcal{C}$ , and statistical length $\mathcal{L}$ as introduced in Result section by Python (version 3.6.1) with numpy library (version 1.12.1).
To calculate them, probability distribution of $a$ and $m$ is necessary.
For case which initial condition is Gaussian, linear Langevin equations follow Gaussian distribution; the mean and covariance of them are sufficient information to describe the distribution \cite{kampen2007} as follows, 
\begin{align}
    \pp{p(a, m)}{t} = \frac{1}{\tau _a}p(a, m) +\frac{1}{\tau _a}[a(t) - \alpha m(t) + \beta l(t)]\pp{p(a, m)}{a(t)} \notag\\  
    + \frac{1}{\tau_m}a(t)\pp{p(a, m)}{m(t)} 
    + T_a\ppt{p(a, m)}{a(t)}+T_m\ppt{p(a, m)}{m(t)}.
  \end{align}
Transition of the mean and the covariance to small time change can be formulated under linear Langevin equations with Gaussian white noise \cite{kampen2007}.
Therefore, the mean and the covariance were calculated by Euler methods with small time change ($\Delta t = 0.0001$).
For initial condition, initial probability of $a$ and $m$ was determined as stationary-state distribution without stimulation ($l = 0$).
$ds^2$ can be calculated from Fisher information matrix and Kullback-Leibler divergence as described below, and these were calculated from the mean and covariance of Gaussian distribution \cite{nielsen2009,amari2016,micali2016} as
\begin{align}
    G(\theta) &= d\mu^T\Sigma^{-1}d\mu+ \frac{1}{2}\mathrm{tr}(\Sigma^{-1}d\Sigma\Sigma^{-1}d\Sigma)\\
    D_\mathrm{KL} (p(a, m;\theta)&|| p(a, m;\theta+\d \theta)) \notag \\
        &= \log \frac{|\Sigma +d\Sigma|}{|\Sigma|}+\mathrm{tr}((\Sigma +d\Sigma)^{-1}\Sigma)\notag \\ 
        &\ + d\mu ^T(\Sigma + d\Sigma )^{-1}d\mu - \mathrm{dim} (p(x))
\end{align}
where $I$ denotes a unit matrix, $\mathrm{dim(\cdot)}$ denotes dimension of a vector, $\Sigma$ is the covariance matrix of $a(t)$ and $m(t)$, and $\mu$ is the mean vector of $a(t)$ and $m(t)$.
In all case, we changed stimulation $l(t)$ from 0 at time 0.
In Fig. \ref{fig2}, we fixed $T_a = 0.005$, and in Fig. \ref{fig3}, we changed $T_a$ at time 0.

\section{Result}
\begin{figure}[h]
    \centering
    \includegraphics[width=\linewidth]{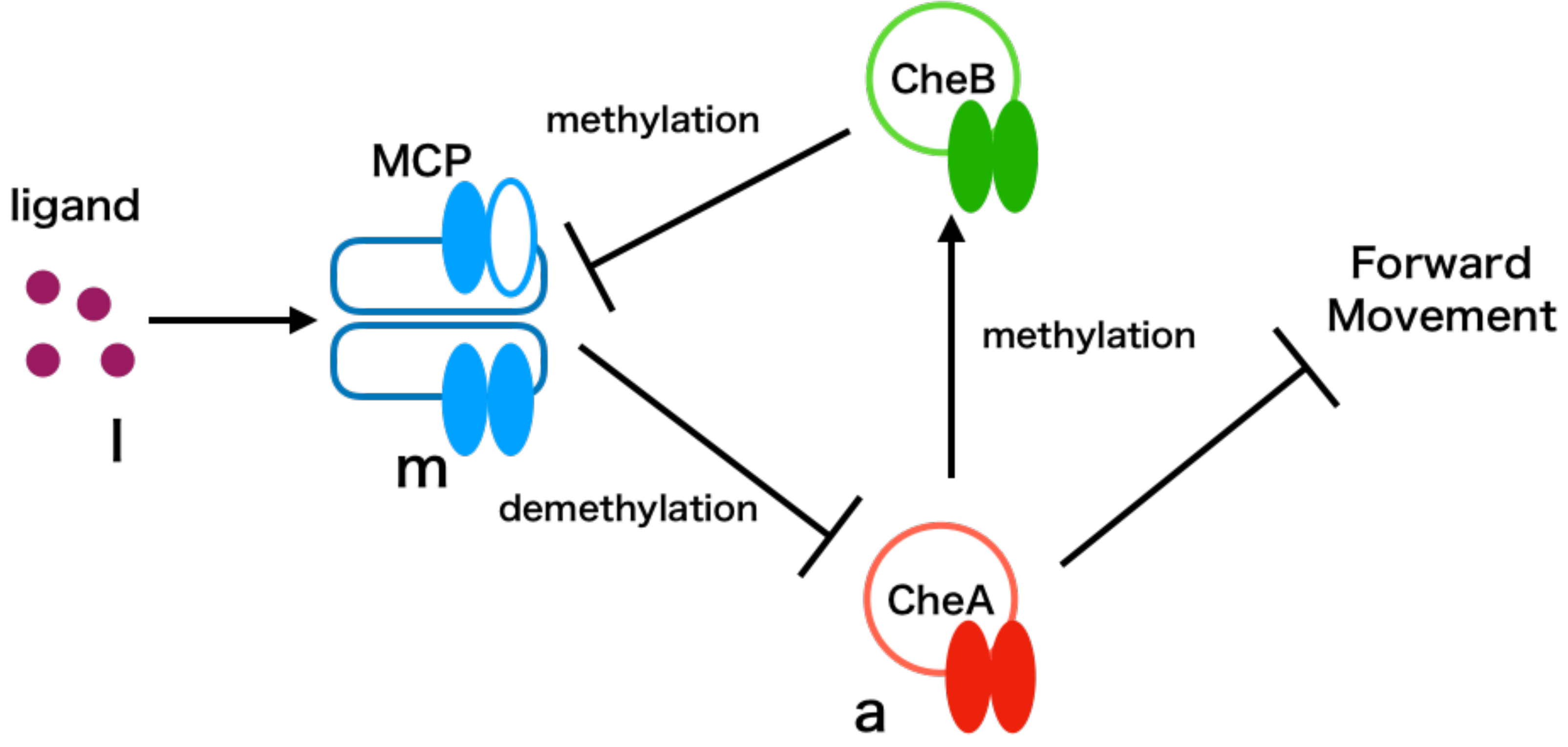}
    \caption{
Schematic model of \textit{E. coli} sensory adaptation.
 External ligand is received by the receptor MCP, and inhibit the kinase CheA autophosphorylation.
CheA phosphorylation inhibits forward movement. 
CheA phosphorylation also phosphorylates CheB, and CheB inhibits MCP by methylation of MCP.
The variables in the model are as follows:
the external ligand level $l(t)$, the CheA activity $a(t)$, and the MCP methylattion level $m(t)$.
}
\label{fig0}
\end{figure}

\begin{figure}[h]
    \centering
    \includegraphics[width=0.7\linewidth]{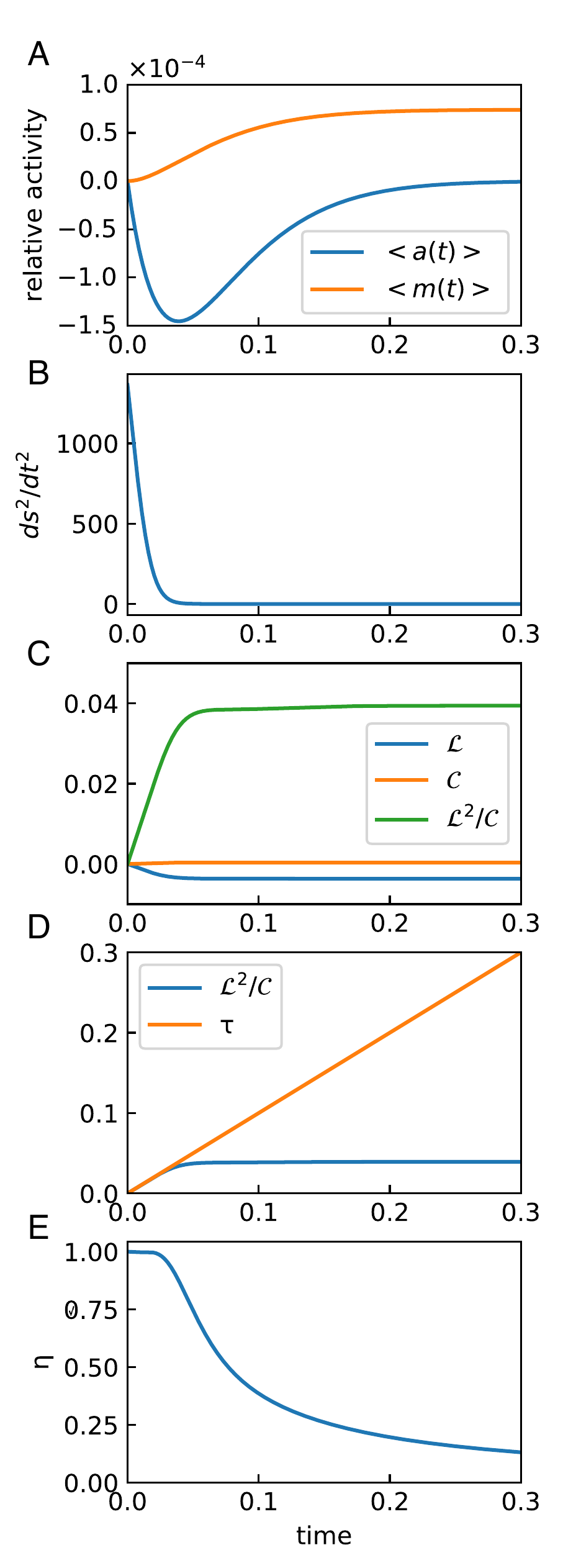}
    \caption{
Numerical simulation of time evolution of squared the time derivative of line element $ds^2/dt^2$ 
    and  the thermodynamic relationship on $\textit{E. coli}$ adaptation model.
Mean activity of the kinase $\ave{a}$ and mean methylation level of the receptor $\ave{m}$ (A) and $ds^2/dt^2$ (B) are shown.
The thermodynamic cost $\mathcal{C}$ , statistical length $\mathcal{L}$ and $\mathcal{L}^2/\mathcal{C}$ are shown in C.
$\mathcal{L}^2/\mathcal{C}$ and $\tau$ are shown in D.
$\tau$ indicates elapsed time from time 0 in Fig. 1.
The inequality $\mathcal{L}^2/\mathcal{C} \geq \tau$ is shown.
The efficiency $\eta$ is shown in E.
}
\label{fig1}
 \end{figure}

\begin{figure}[h]
    \centering
    \includegraphics[width=\linewidth]{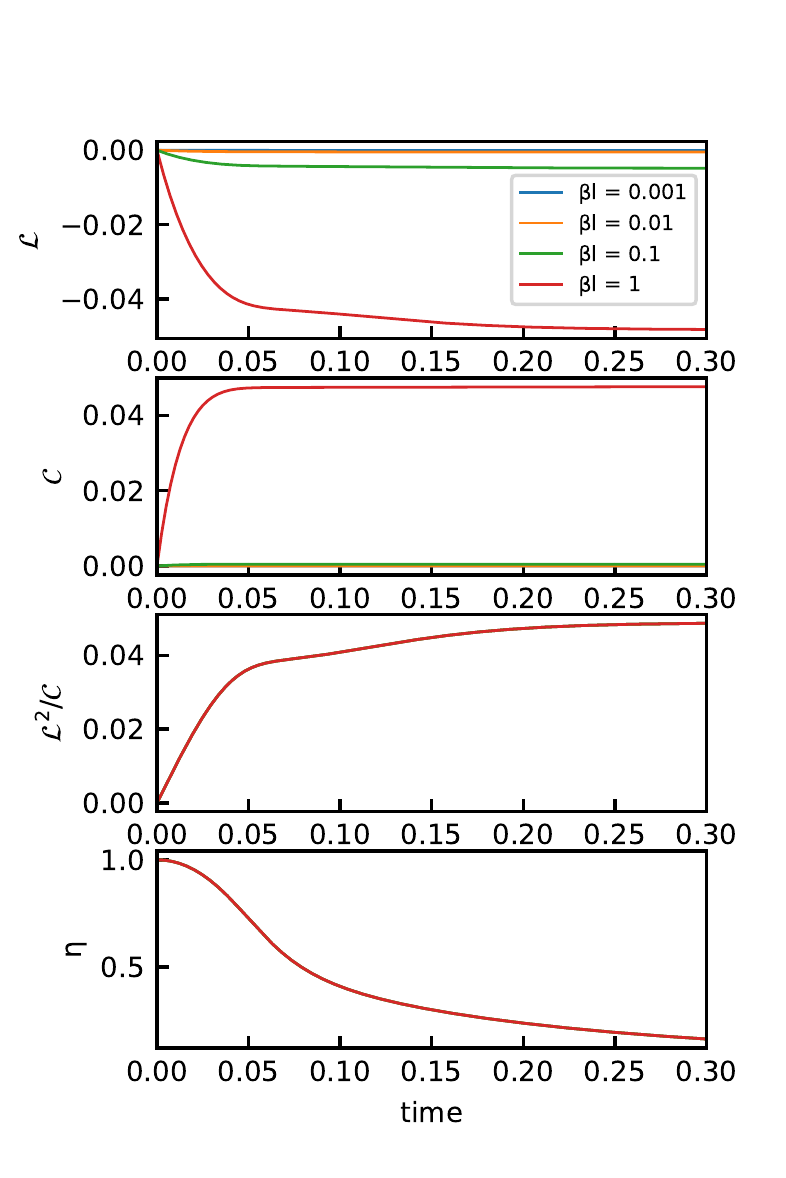}
    \caption{
Numerical simulation of the relationship between the speed of the adaptation and the thermodynamic cost on $\textit{E. coli}$ adaptation model with different stimulation level. 
The thermodynamic cost $\mathcal{C}$ (top), the statistical length $\mathcal{L}$ (top middle), $\mathcal{L}^2/\mathcal{C}$ (bottom middle) and the efficiency $\eta$ (bottom) are shown.
 In all case, $\mathcal{L}^2/\mathcal{C}$ and $\eta$ are almost same values.
}
\label{fig2}
 \end{figure}

\begin{figure}[h]
    \centering
    \includegraphics[width=\linewidth]{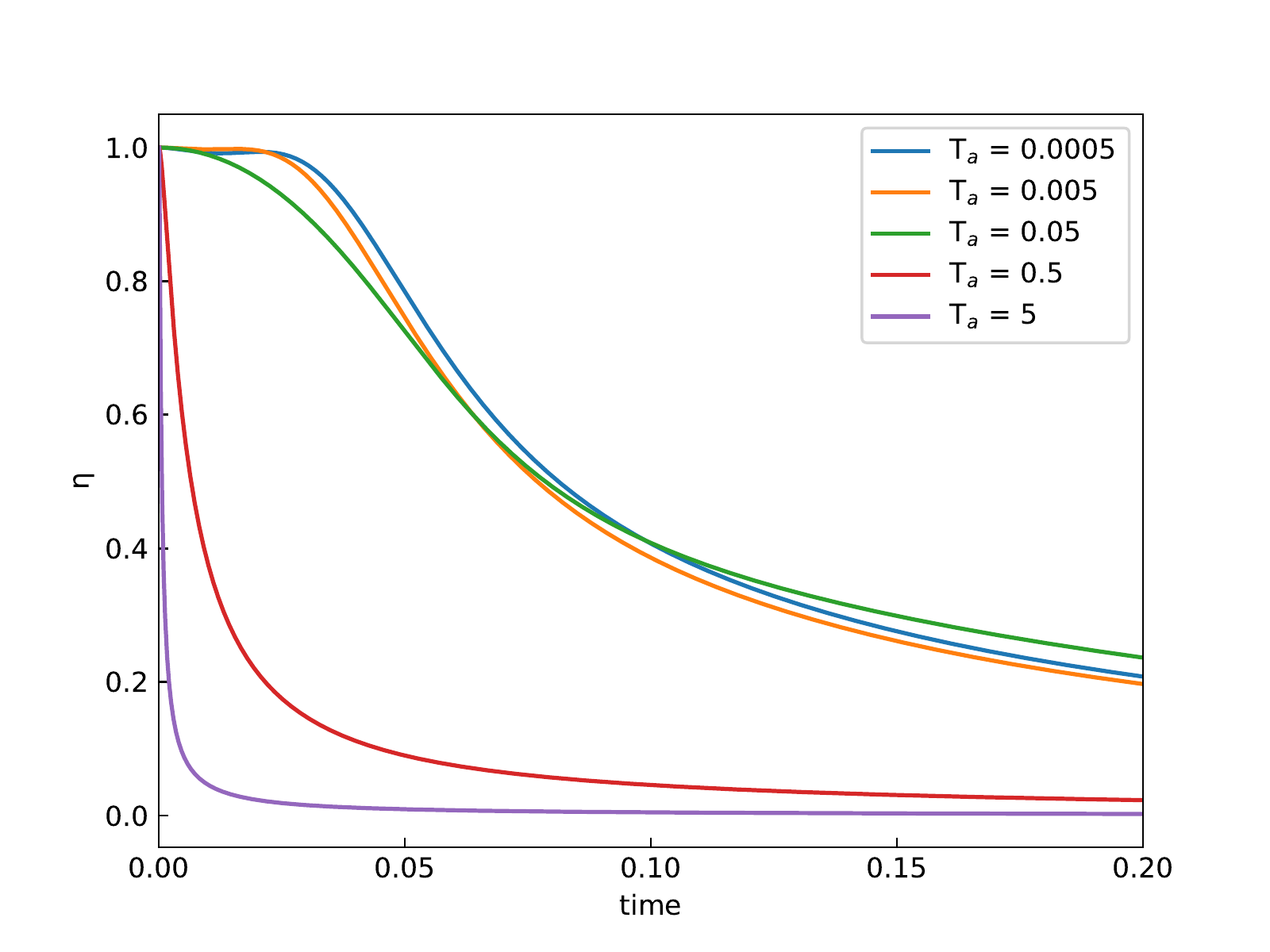}
    \caption{
Numerical simulation of the efficiency $\eta$ on $\textit{E. coli}$ adaptation model with different noise level.
}
\label{fig3}
 \end{figure}

\subsection{Information geometry in \textit{E. coli} sensory adaptation model}
In \textit{E. coli}, sensory adaptation for chemotaxis has been described in detail, and 
 an external ligand is received by a receptor, and inhibit a kinase phosphorylation (Fig. \ref{fig0}) \cite{hazelbauer2008,micali2016,bi2018}.
As previous research, mathematical model of the adaptation was formulated as linear Langevin equations (Eqs. \ref{eq:a} and \ref{eq:m}) \cite{tu2008, sartori2011, ito2015}.
This model is a coarse-grained model, which is enough simple for thermodynamic analysis, and
$a(t)$ expresses the kinase activity corresponding to CheA, and $m(t)$ expresses the metylation level of the receptor MCP.
As described above, an external signal inhibits the kinase activity, and methylates the receptor (Fig. \ref{fig1}A).
The levels are random variables because they fluctuate by noise; therefore, the levels can be expressed by probability distributions.
The probability of $a(t)$ and $m(t)$ follows Gaussian distribution under a condition that initial distribution is Gaussian for linear Langevin equation.

Next, we applied information geometry to this model.
In information geometry, we can measure local distance between two different distribution on a statistical manifold \cite{amari2016}.
For exponential family, such as Gaussian distribution, a square of an infinitesimal distance $ds$ between $p(x; \theta)$ and $p(x; \theta + d\theta)$ is defined as 
\begin{align}
ds^2 &= 2\KL{p(x;\theta)||p(x;\theta+d\theta)}\label{eq:dskl}\\
 &= \sum g_{ij}(\theta)d\theta_i d\theta_j
\end{align}
where $\theta$ is a parameter which expresses the distribution, $\mathrm{D_{KL}}$ indicates Kullback-Leibler divergence,
and $G(\theta) = (g_{ij}(\theta))$ is Fisher information matrix \cite{amari2016}.
As noted above, in this model, the probability of $a(t)$ and $m(t)$ is according to Gaussian distribution when initial distribution is Gaussian.
Therefore, initial probability distribution of $a(t)$ and $m(t)$ must be a Gaussian as an assumption.
By this assumption, the distribution of $a(t)$ and $m(t)$ is considered as exponential family.

As previous research \cite{ito2017,ito2018}, we consider the value $ds^2/dt^2$ as
\begin{align}
\frac{ds^2}{dt^2} &= \left(\frac{\sqrt{ds^2}}{dt} \right)^2\\
 &= \left(\frac{1}{2\sqrt{ds^2}}\frac{ds^2}{dt}\right)^2.
 \end{align}
 We noted that $ds^2/dt^2 = \sum g_{i,j} d\theta _i/dt d\theta _j/dt$.

Next, the statistical length $\mathcal{L}$ from time 0 to $\tau$ is defined as $\mathcal{L} = \int ds = \int ds/dt dt$ as previous research \cite{wootters1981,ito2017}.
Therefore,
\begin{align}
\mathcal{L} &= \int_0^\tau \frac{ds}{dt}dt.
 \end{align}
 Moreover, thermodynamic cost $\mathcal{C}$ is defined as
\begin{align}
\mathcal{C} &= \int_0^\tau \left(\frac{ds}{dt}\right)^2dt.
 \end{align}
As described in previous research \cite{ito2018},
this quantity $\mathcal{C}$ is related to entropy change of the system $\sigma_{sys}$.
Time derivative of entropy change of the system for transition $-\partial \sigma_{sys}/\partial t$ is equivalent to the thermodynamic cost $\mathcal{C}$.
From Cauchy-Schwarz inequality \cite{crooks2007},
a relationship between the speed of the adaptation and the thermodynamic cost is 
\begin{align}
\int_0^\tau dt \int_0^\tau \left(\frac{ds}{dt}\right)^2dt &\geq \left(\int_0^\tau \frac{ds}{dt}dt\right)^2\\
 \tau &\geq \frac{\mathcal{L}^2}{\mathcal{C}}
 \end{align}
 according to the previous studies \cite{ito2017,ito2018}.
 This inequality expresses the lower bound of the state transition of $a(t)$ and $m(t)$.
 The state transition means dynamics of the adaptation; this inequality gives the bound for the adaptation speed.
 The equality holds if speed of $ds^2/dt^2$ is independent from time.
 Moreover, the efficiency of the transition $\eta$ is defined as
\begin{align}
 \eta = \frac{\mathcal{L}^2}{\tau \mathcal{C}}\\
0 \leq \eta \leq 1.
 \end{align}
 When $\eta$ is 1, the transition is most efficient.

Next, we numerically calculated the values described above (Fig. \ref{fig1}).
 Numerical simulation showed that $ds^2/dt^2 \geq 0$.
 This property is due to the non-negativity of the squared value.
 Moreover, the relationship between the speed of the adaptation and the thermodynamic cost gives tight bound to transition time $\tau$ (Fig. \ref{fig1} D).
From these results, we showed the relationship with numerical simulation. 
As shown in Fig. \ref{fig1} E, there are three phases for $\mathcal{L}^2/\mathcal{C}$ and $\eta$ dynamics: \textasciitilde0.02 (respond to ligand), \textasciitilde0.05 (adapting), and 0.05\textasciitilde  (completion of adaptation).
The first phase is almost efficient as the thermodynamic boundary (the efficiency $\eta$ is almost 1).
After starting adaptation, the transition became inefficient.
After achieving adaptation, $\mathcal{L}^2/\mathcal{C}$ became stable because $a(t)$ and $m(t)$ are under stationary state.
Next, stimulation levels were changed (Fig. \ref{fig2}).
The thermodynamic cost $\mathcal{C}$ and statistical length $\mathcal{L}$ were changed by stimulation level,
but the efficiency was not changed.
Finally, we changed the noise levels of stimulation (Fig. \ref{fig3}).
Intuitively, small noise makes high efficiency.
We obtained similar tendency to intuition, but for small noise, this is not correct.
For the efficiency of first phase as describe above, the efficiency at noise level $T_a = 0.0005$ is higher than one at noise level $T_a = 0.005$.
For the second phase, the efficiency at $T_a = 0.0005$ is worse than one at $T_a = 0.005$.
At $T_a = 0.00005$, the efficiency became worse.
These results indicate that there is the best noise level to achieve the adaptation.
These results firstly demonstrated that the efficiency of the adaptation speed is highly robust for ligand strength, but drastically changed for noise level of ligand on \textit{E. coli} adaptation.

\section{Discussion}
Introducing stochastic thermodynamics and information geometry, we succeeded in defining the relationship between the speed of the adaptation and the thermodynamic cost in sensory adaptation model of \textit{E. coli}.
We assumed that the initial distribution is Gaussian and Langevin equations are linear, but the scope of application is still wide.
 In previous research \cite{ito2017}, only master equation model is discussed, 
 but the Langevin equation, which we applied is also used for numerous models including osmotic sensing in yeast, olfactory sensing in neurons and light sensing \cite{nakatani1991,menini1999,hohmann2002,hazelbauer2008,lan2012,shidara2017}.
Therefore, our quantification can be used for various systems in biology.

We introduce stochastic thermodynamics and information geometry to \textit{E. coli} sensory adaptation model.
This model has been investigated well by stochastic thermodynamic approach \cite{sartori2011, lan2012, sartori2014, ito2015, ito2017}.
However, the speed of the adaptation has been less discussed than the thermodynamic cost and robustness of the adaptation.
These factors are important components for the adaptation; however, the speed of the adaptation is also important factor because the adaptation speed changes by repetitive stimulation \cite{chalasani2007,shidara2017}.
Previous research discussed the speed of the adaptation by energy-speed-accuracy relationship \cite{lan2012}.
Using the formulation, we can discuss the relationship between time constant of the models and the thermodynamic cost.
On the other hand, using our formulation, we can discuss the efficiency over time as shown in Figs. \ref{fig2} and \ref{fig3}.
Therefore, the relationship described in this work is crucial for understanding adaptation quantitatively.

The efficiency of the adaptation did not show proportional relationship against the noise level (Fig. \ref{fig3}), and seems to have appropriate noise level for efficient adaptation.
In the context of signal detection, small noise improves detection performance, and this is known as stochastic resonance \cite{luca1998,mark2009}.
This phenomenon is similar as our observation (Fig. \ref{fig3}).
There is no research directly related to our result. However, stochastic resonance has been reported in some types of adaptation \cite{stocks2000,mark2009,tutu2011};
therefore, our result might be new type of the stochastic resonance.

We showed that the efficiency of the adaptation speed with different noise levels and different stimulation strengths.
For stimulation level, the efficiency is highly robust, but drastically changed for noise level.
In olfactory system in \textit{Caernohabditis elegans}, the adaptation speed of a second messenger (cGMP) is highly robust for different stimulation strengths \cite{shidara2017}.
Therefore, our result is coincident with biological experiment.
For the adaptation speed, although the time constant is usually measured, 
 it is difficult for further quantification in experimental study \cite{lan2012}.
Moreover, quantifying the efficiency of the adaptation speed is hardly difficult by experiments only.
Using optogenetics, stimulation strength and noise level can be controlled \cite{gepner2015,gorur2017,chen2018}.
Therefore, our result can be applied to not only theoretical models, but also experimental research, and provides quantitative understanding for adaptation mechanism.

\section*{Author Contributions}
K.A designed and performed the research; K.A. wrote the original draft of paper, and K.A. and K.O reviewed and edited the paper. K.O supervised the work.

\section*{Acknowledgement}
We thank for Dr. Kohji Hotta at Keio University for discussion.

\section*{Conflict of Interest}
The authors declare no conflicts of interest associated with this manuscript.

\section*{Funding}
This research did not receive any specific grant from funding agencies in the public, commercial, or not-for-profit sectors.

\bibliography{ref}
\end{document}